\documentclass[twocolumn,showpacs,preprintnumbers,amsmath,amssymb]{revtex4}

\usepackage{graphicx}
\usepackage{dcolumn}
\usepackage{bm}

\begin{document}

\preprint{A FOCUS Preprint}

\title{Extracting quantum dynamics from genetic learning algorithms\\ through principal control analysis}

\author{J. L. White, B. J. Pearson, and P. H. Bucksbaum}
 \affiliation{FOCUS Center, Physics Department, University of Michigan.}

\date{\today}

\begin{abstract}
Genetic learning algorithms are widely used to control ultrafast
optical pulse shapes for photo-induced quantum control of atoms and
molecules. An unresolved issue is how to use the solutions found by
these algorithms to learn about the system's quantum dynamics. We
propose a simple method based on covariance analysis of the control
space, which can reveal the degrees of freedom in the effective
control Hamiltonian. We have applied this technique to stimulated
Raman scattering in liquid methanol. A simple model of two-mode
stimulated Raman scattering is consistent with the results.
\end{abstract}

\maketitle

The central challenge of coherent control of quantum dynamics is
to find the optimal path to guide a quantum system from its
initial state to some target final
state\cite{TannorRice88,BrumerShapiro92}. Several theoretical
methods have been developed to aid this search
\cite{TannorKosloffRice86,RabitzOptCon88}, and there has been
considerable experimental success as
well\cite{ShapiroExpt96,GordonHIScience95}.  However, in all but
the simplest systems, the search is hampered by incomplete
knowledge of the system Hamiltonian. Strongly coupled systems such
as large molecules in condensed phase are so complicated that it
is nearly impossible to calculate optimal pulse shapes in advance.

Feedback learning algorithms overcome this limitation by using the
physical system itself to explore its own quantum dynamics through
an experimental search \cite{JudsonRabitz92}. A typical search
experiment compares the ability of several thousand different shaped
laser pulses to transform the system $|\psi\rangle$ from its initial
state at time $t=0$ to some desired target state $|\chi\rangle$ at a
later target time $t=T$.  Examples of transformations that have been
studied include molecular photodissociation, atomic photoexcitation
and photoionization. The pulse shapes are selected through a
fitness-directed search protocol, such as a genetic algorithm
\cite{HollandSciAm92}.  The \textit{fitness} is a measured quantity
proportional to the \textit{objective functional} $J[H;x_{i}]$,
which is the square of the projection of $|\chi\rangle$ onto
$|\psi\rangle$ at the end of the experiment:
\begin{equation}\label{eq:Jfunction}
J[H;x_{i}]=|\langle\chi (T)|\psi (T)\rangle|^{2}.
\end{equation}
$J$ depends on the Hamiltonian $H$ for the system evolution, which
depends in turn on the laser electric field $E(t)$ determined by the
settings of the $n$ pulse shape control parameters $x_{i}$,
$i=1\ldots n$. $J$ reaches its extreme value for the optimal pulse.
This pulse can be calculated using optimal control theory if $H$ is
known \cite{RabitzOptCon88}; otherwise, it must be discovered
through the learning search.

Several recent papers have suggested modifications or extensions of
learning feedback that can measure properties of the system
Hamiltonian
\cite{GeremiaRabitz01,Mitra03,WosteScience03,GeremiaRabitz02}. Here
we propose a different approach based on analysis of the trial
experiments.  We will show that the ensemble of trial pulse shapes
can reveal essential features of the dynamics.

Genetic algorithms and similar evolutionary search strategies have
many different variations \cite{GenHandbk91}. Our implementation
starts with approximately 50 randomly generated optical pulse shapes
produced by spectrally filtering an ultrafast laser pulse
\cite{Warren}. Each pulse shape is described by a column matrix of
control parameters $x_{i}$ called a \textit{genome} consisting of
about 25 numbers (\textit{genes}), each encoding the amplitude
and/or phase of a different segment of the optical spectrum.  The
control target is measured for each pulse shape. Then the algorithm
creates a new \textit{generation} of pulse shapes by combining
attributes of the fittest members of the previous generation
\cite{PearsonPRA01}. After several generations, the pulse shapes
usually cluster near high fitness regions of the search space. When
the algorithm finds a pulse shape or several shapes that cannot be
improved over many generations, the search stops, and the highest
fitness pulse shape is declared the optimal solution to the search.
We test 1000 to 10,000 pulse shapes in a typical experiment. We
maintain a record of every pulse shape, its fitness, and its
parentage \textit{(genealogy)}.

The learning algorithm achieves control without prior knowledge of
the system Hamiltonian, and has far more degrees of freedom $n$ than
the minimum required for control. The number of possible solutions
is exponential in $n$.  In a typical search, the phase of each color
is adjusted by the spectral phase filter to a precision of about ten
degrees, so there are $2^{5}$ possible values of each gene. This
means that the number of possible solutions for a genome of length
25 is $2^{5 \times 25}\simeq 4 \times 10^{37}$. Genetic algorithms
can search this large state space with great
efficiency\cite{HollandSciAm92}. Unfortunately, simply finding a
good solution has not often provided significant insight into the
system dynamics or Hamiltonian. The optimal pulse shape found by the
learning algorithm, while sufficient to achieve control, is often
complicated and may contain unnecessary features.

The conditions for reaching an extremum in $J[H;x_{i}]$ may only
depend on two or three essential features of the control field
$E(t)$. These features are not obvious in the successful genome
because they may depend on all 25 genes. The Hamiltonian could be
written in a much simpler form if these essential degrees of freedom
($u_{j}$) were found. Here we show how to establish the $u_{j}$
through covariance analysis of the pulse shapes evaluated during the
learning search\cite{PCA_Jolliffe02}. Covariance analysis is
commonly used to reduce the dimensionality of and to find patterns
in high dimensional data sets.

We propose to apply these techniques, not to sets of data, but to
the control space for the experiment. Linear combinations of genes
with high fitness should appear correlated in the fitness-driven
genetic algorithm. These correlated linear combinations correspond
to the principal components of the control field that direct the
quantum dynamics under investigation. \textit{Principal control
analysis} is the application of covariance techniques to a fitness
directed search.

Principal control analysis is implemented on our system by
calculating the covariance matrix of the set of all pulse shapes in
the search, defined by:
\begin{equation}\label{eq:statistics}
    C_{ij}=\langle \delta_{i}\delta_{j}\rangle-\langle \delta_{i} \rangle\langle \delta_{j} \rangle,
\end{equation}
where the expressions $\delta_{i} = x_{i+1} - x_{i}, i = 1 \ldots
n-1$ are the nearest neighbor phase differences. By using the phase
differences in the analysis we remove the ambiguity associated with
the unimportant global phase. The covariance matrix is not the only
measure of correlation. We could also weight the terms of the
covariance matrix using the fitness, or normalize each term to the
individual gene variance. In this paper, we will use the simple
covariance. This is appropriate because all the $\delta_{i}$'s are
of the same type.

Once the covariance matrix is determined, we calculate its
eigenvectors and eigenvalues. Each eigenvalue $\lambda_{j}$ measures
the variance of the projections onto the corresponding eigenvector.
This has a special meaning for a learning control search: it shows
how far the control setting moved during the learning process. A
small subset of eigenvalues usually contains most of the weight of
the trace of the covariance matrix. The correlations of the
projections with the pulse shape fitnesses allow us to determine
which control directions were most important for the physical
process under consideration. The controls expressed in the basis of
the eigenvectors are uncorrelated: Each of these controls changes
the fitness without correlation with the others over the search set.

We propose that these eigenvectors with the largest fitness
correlation are the essential control directions ($u_{j}$). We
expect the eigenvectors with the larger eigenvalues to be most
strongly correlated with the fitness of a pulse shape solution.
Conversely, a low correlation indicates those eigenvectors that have
not contributed substantially to increasing the fitness during the
search. These eigenvectors correspond to extraneous dimensions,
which could be eliminated (i.e., their projection set to zero)
without losing substantial control.

By projecting the GA solutions onto the $k<n$ eigenvectors that
correlate best with the fitness (the principal controls), we reduce
the dimension of the control space.  The solutions with highest
fitness, when expressed in the reduced basis of the principal
controls, represent the essential features of the search solutions.
The objective functional also takes on a simpler form in this basis:
\begin{equation}\label{newJ}
    J=J[H;u_{1,\ldots,k}]
\end{equation}
where each $u_{j}$ can assume a range of values on the order of
$\pm\sqrt{\lambda_{j}}$. The search for the specific target state
is now a matter of optimizing each control $u_{j}$ over this
range.

In summary, we propose to apply covariance techniques to the control
space of learning feedback experiments.  Control values derived from
the genomes are analyzed by a covariance matrix, defined in Eq.
\ref{eq:statistics}. The matrix eigenvectors are independent control
directions. The correlation of the fitness with the eigenvectors
suggests which control directions are the most important. The
corresponding eigenvalues indicate the necessary excursion along
each axis. Therefore, searches conducted in the eigenvector basis
are more efficient. Finally, the best solutions are found by
projecting the optimal learning control solutions onto the important
control directions.

\begin{figure}[h]
\includegraphics[width=8.5cm,height=12.0cm]{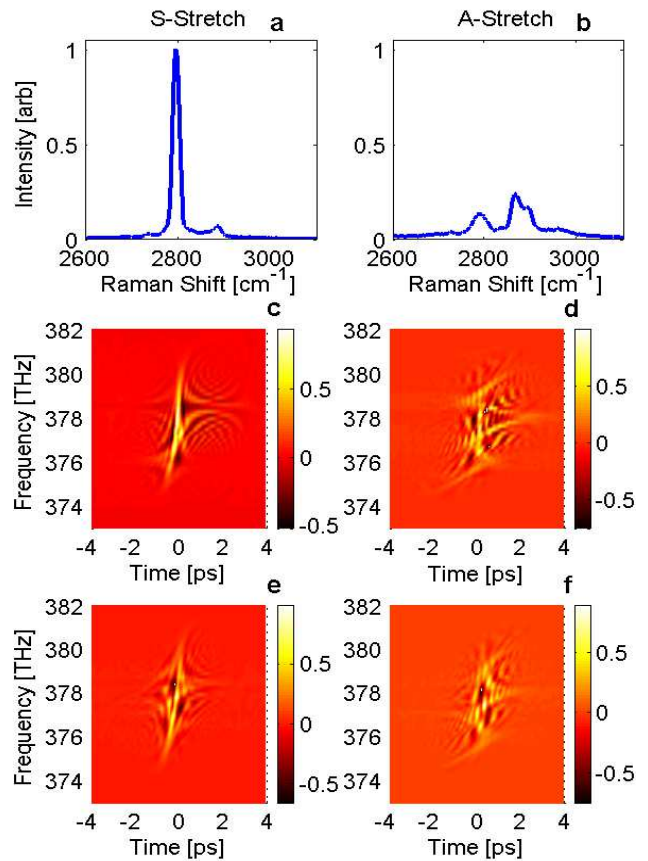}\\
\caption{Left: Raman spectra (a) and Wigner representation (c) of
the optimal pulse shape after optimizing the symmetric C-H stretch
mode. Wigner representation (e) of the control pulse shape found by
principal control analysis. Right: Similar for antisymmetric
stretch.}\label{fig:meohWigner}
\end{figure}

We now apply this analysis to a well-studied control problem:  the
selective excitation of vibrational modes in liquid methanol.  The
experiment has been described previously
\cite{Pearson2,PearsonFemto,PearsonPRA01}.  An intense shaped 800 nm
ultrafast laser pulse (the pump laser) is focused into a cell
containing methanol.  Above a threshold fluence, the pump induces
stimulated Raman scattering into either the symmetric or
antisymmetric Raman-active C-H stretch mode.  Either mode can be
selectively excited by adjusting the shape of the pump pulse through
phase shaping and/or amplitude shaping of its spectrum.

Figure \ref{fig:meohWigner} depicts the results of a phase-only
feedback control experiment. Panel a (b) shows the Raman spectra
produced by the optimal pulse for the symmetric (antisymmetric)
stretch mode. This learning search included $2720$ different pulse
shapes. Moderate fitness increases were observed for either target
mode after $25$ generations.  The fitness increase was greater for
the symmetric mode. This is typical of our searches based on
phase-only control\cite{PearsonFemto}.

A Wigner representation of the pulse shape solution found by the
learning algorithm is plotted in panel c (d). The Wigner function is
a spectrally resolved field auto-correlation:
\begin{equation}\label{Wigner}
    W(\omega,t)=\int d\omega'E(\omega-\omega')E^{*}(\omega+\omega')e^{-2i\omega ' t}
\end{equation}
Wigner representations are complete time-frequency spectrograms of
the optical control field (up to a global phase), but the important
features leading to control of the methanol are obscure. The
inability to interpret the result easily is typical of many GA
search solutions \cite{LevisScience01,GerberLiqPhase01}.

\begin{figure}[h]
\includegraphics[width=8.5cm,height=12.0cm]{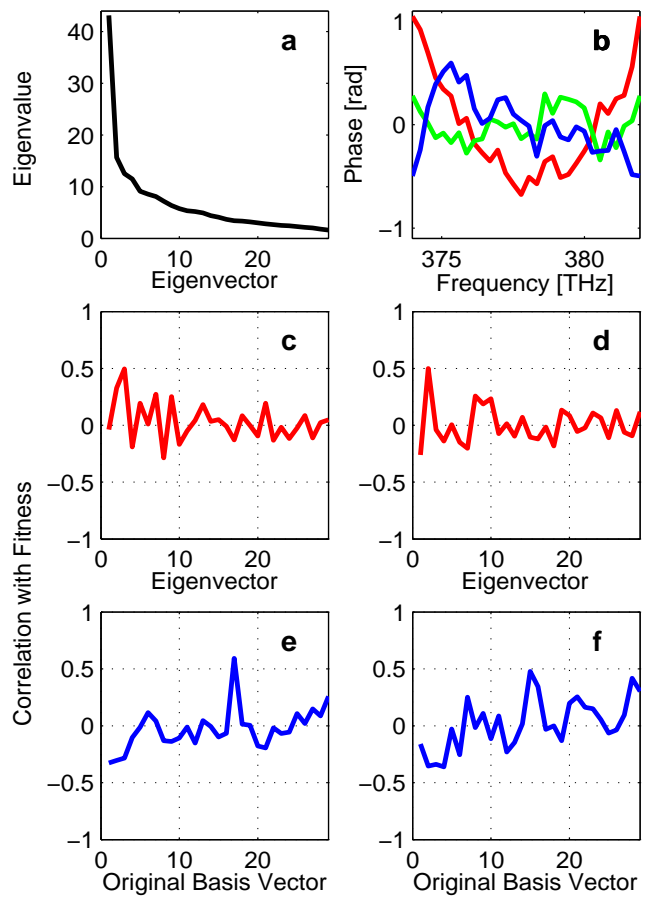}\\
\caption{Panel a: Eigenvalues of the covariance matrix in descending
order. Panel b: Phase functions associated with the three principal
control eigenvectors. For the symmetric stretch, the principal
controls correspond to the third (blue) and second (red)
eigenvalues. Antisymmetric stretch corresponds to the second (red)
and first (green) eigenvalues. Panel c (d): Correlation of fitness f
with the control vectors for the symmetric (antisymmetric) stretch
ordered by eigenvalue. Specifically, the correlation is
$(<\eta_{i}f>-<\eta_{i}><f>)/\sigma_{\eta_{i}}\sigma_{f}$. Panel e
(f): The correlation in the original basis ordered by frequency:
$(<\delta_{i}f>-<\delta_{i}><f>)/\sigma_{\delta_{i}}
\sigma_{f}$.}\label{fig:fitcorr}
\end{figure}

The principal control analysis of this problem begins with a single
covariance matrix (Eq. \ref{eq:statistics}) for the entire
population of pulse shapes evaluated in the two feedback
experiments. The physical system under control was the same in the
two problems; only the target was different. We therefore expect the
independent searches to be nearly spanned by a small number of
eigenvector controls. The covariance matrix is not simple to
interpret because the principal components in this problem are
widely distributed among all of the control settings in the search
space. However, the essential features of the control problem begin
to emerge if the covariance matrix is diagonalized. This can be seen
in Fig. \ref{fig:fitcorr}a, where the eigenvalues of the covariance
matrix are plotted in descending order. The algorithm has made large
excursions only along the few control directions that have large
eigenvalues. The phase functions associated with the three largest
eigenvalues are shown in Fig. \ref{fig:fitcorr}b. These control
directions are also the most strongly correlated with fitness.
Figure \ref{fig:fitcorr}c (d) plots the correlation of eigenvector
with fitness for the symmetric (antisymmetric) stretch. The three
most significant eigenvectors are the same as the control directions
that correlate most strongly with the fitness. Therefore we propose
that the dimension of the search space can be reduced without
inhibiting control.

Each pulse shape can now be re-expressed in the eigenvector basis.
To arrive at the essential features of the best pulse, we calculate
the projections $\eta_{k}$ of the optimal pulse shape onto the
principal control directions $u_{k}$. When these $u_{k}$ are
expressed in the original basis, their components are individual
discrete frequencies that make up the field. Therefore,
$\sum_{j=1}^{k} \eta_{j}u_{j}(\omega)$ produces a pulse that
contains traits necessary to achieve the target, with minimal
extraneous features. Figure \ref{fig:meohWigner}e (f) shows the
Wigner plot of this essential pulse for the symmetric
(antisymmetric) stretch mode. The essential pulse for the symmetric
stretch preserves $66\%$ of the original pulse shape vector
($\sum_{j=1}^{k} \eta_{j}^{2}(\omega)=66\%$). For the antisymmetric
stretch, this number is $88\%$.

\begin{figure}[h]
\includegraphics[width=8.5cm,height=8.5cm]{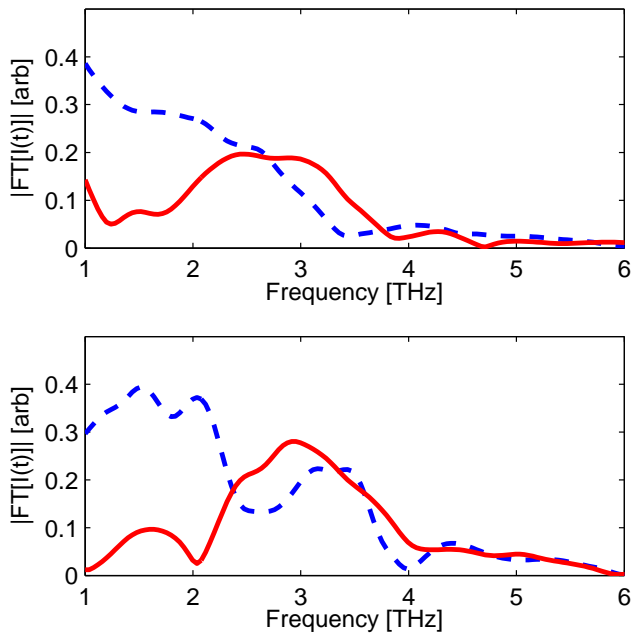}\\
\caption{Top: Magnitude of the Fourier transform of $I(t)$ for the
optimal pulse shape found by the learning algorithm (dashed) and for
the essential pulse shape found through principal control analysis
(solid) for the symmetric stretch mode. Bottom: Similar for the
antisymmetric stretch mode.}\label{fig:fourier}
\end{figure}

This procedure is independent of the specific nature of the
physical system or the dynamical Hamiltonian; however, the
principal control directions can be used to construct a simplified
interaction Hamiltonian, since the laser electric field now only
depends on a few parameters:
\begin{equation}\label{eq:sepH2}
    H(t)=H(E[u_{j=1\ldots k};t]).
\end{equation}
This can provide important constraints.  For example, a recent paper
demonstrated a control mechanism for SRS in methanol based on
periodicity in $I(t)$ \cite{Pearson2}. The Fourier transform of
$I(t)$ then reveals the most important Raman coupling frequencies in
the problem \cite{SilberbergNature98}. Figure \ref{fig:fourier}
(top) shows the magnitude of the $FT[I(t)]$ for the optimal pulse
shape found by the learning algorithm for the symmetric stretch mode
(dashed) and for the same pulse shape projected onto the principal
control directions $u_k$ (solid). Figure \ref{fig:fourier} (bottom)
shows similar plots for the antisymmetric mode.

For both modes, projecting the optimal pulses onto the principal
control directions enhances the frequency components around $3$ THz.
The coupling frequency found through principal control analysis
agrees with the model based on the mode separation in methanol.
Additionally, comparing the phases of the Fourier transforms for
each mode yields a phase difference of $\pi/2$ in the region around
$3$ THz, consistent with the model described in reference
\cite{Pearson2}.

In conclusion, we have shown how covariance analysis of a genetic
search algorithm can uncover essential features of the dynamical
Hamiltonian. Although our example involved only phase-shaping of the
optical field, this technique should be applicable to any system
where fitness-directed learning algorithms have been used to reveal
the path from an initial quantum state to a target. Principal
control analysis can also be incorporated into the experimental
search protocol. By discovering the principal controls, it should be
possible to search the space more efficiently, and to test ideas
about the system dynamics as the search is proceeding. The method
should be most useful in cases where the dynamics can be described
by only a few principal degrees of freedom, which are linear
combinations of the control parameters of the search space.

We thank Anna Amirdjanova, Jayson Cohen, Vladimir Dergachev, Chitra
Rangan, Ben Recht, and Tom Weinacht for valuable discussions.  This
work was supported by the National Science Foundation under grant
9987916.

\bibliography{pca_23}

\end{document}